Attached are 7 files, separated by lines of the form:
------------------------------------------------------------------------------
The attached files are:

main.tex
aps.sty
aps10.sty
preprint.sty
revtex.sty
fmextrpf.ps
faextrpf.ps

The Postscript files are not automatically included in the Latex and need to
be printed separtately.

------------------------------------------------------------------------------
\documentstyle[preprint,revtex]{aps}
\begin{document}
\draft
\begin{title}
Meson Decay Constant Predictions of the Valence Approximation\\ to Lattice QCD
\end{title}
\author{F. Butler, H. Chen, J. Sexton\cite{Trinity}, A. Vaccarino,\\
and D. Weingarten}
\begin{instit}
IBM Research \\
P.O. Box 218, Yorktown Heights, NY 10598
\end{instit}
\begin{abstract}
We evaluate $f_{\pi}/ m_{\rho}$, $f_K/ m_{\rho}$, $1/f_{\rho}$, and $
m_{\phi}/(f_{\phi} m_{\rho})$, extrapolated to physical quark mass, zero
lattice spacing and infinite volume, for lattice QCD with Wilson quarks
in the valence (quenched) approximation.  The predicted
$m_{\phi}/(f_{\phi} m_{\rho})$ differs from experiment by less than its
statistical uncertainty of approximately 15\%.  The other three
constants are 10\% to 20\% below experiment, equivalent to between one
and two times the corresponding statistical uncertainties.
\end{abstract}
\pacs{}
\narrowtext

In a recent paper~\cite{Butler93} we presented lattice QCD predictions
for the masses of eight low-lying hardrons, extrapolated to physical
quark mass, zero lattice spacing, and infinite volume, using Wilson
quarks in the valence (quenched) approximation. The masses we found were
within 6\% of experiment, and all differences between prediction and
experiment were consistent with the calculation's statistical
uncertainty.  We argued that this result could be interpreted as
quantitative confirmation of the low-lying mass predictions both of QCD
and of the valence approximation. It appeared to us unlikely that eight
different valence approximation masses would agree with experiment yet
differ significantly from QCD's predictions including the full effect of
quark-antiquark vacuum polarization.

We have now evaluated the infinite volume, zero lattice spacing,
physical quark mass limit of $f_{\pi}/ m_{\rho}$, $f_K/ m_{\rho}$,
$1/f_{\rho}$, and $m_{\phi}/(f_{\phi} m_{\rho})$.  To our knowledge
there have been no previous systematic calculations of this physical
limit of lattice meson decay constants. A review of earlier work is
given in Ref. \cite{Toussaint}.  The predicted $m_{\phi}/(f_{\phi}
m_{\rho})$ differs from experiment by less than its statistical
uncertainty of approximately 15\%.  The other three constants are 10\%
to 20\% below experiment, equivalent to between one and two times the
corresponding statistical uncertainties.

The decay constant predictions we obtain may appear to be marginally in
conflict with our earlier mass results.  A simple argument suggests,
however, that valence approximation decay constants actually should lie
at least some amount below the predictions of the full theory and
therefore, presumably, below experiment
\cite{Fermilab}.  The valence approximation may be viewed as
replacing the momentum and frequency dependent color dielectric constant
arising from quark-antiquark vacuum polarization with it zero-momentum
limit \cite{Weingar81}.  The valence approximation might thus be
expected to be fairly reliable for low-lying baryon and meson masses,
which are determined largely by the low momentum behavior of the
chromoelectric field.  The effective quark charge used in the valence
approximation may be thought of as the product of the zero-momentum
dielectric constant and the quark's true charge.  The valence
approximation's effective quark charge at higher momenta can then be
obtained from the low momentum charge by the Callan-Symanzik equation.
The leading coefficient $\beta_0$ entering the Callan-Symanzik equation
in the valence approximation is larger than $\beta_0$ of the full
theory, which is reduced by dynamical quark vacuum polarization. It
follows that the quark charge in the valence approximation will fall
faster with momentum than it does in the full theory.  At sufficiently
short distance the attractive quark-antiquark potential in the valence
approximation will therefore be weaker than in the full theory. Meson
wave functions in the valence approximation will then be pulled into the
origin less than in the full theory and will be smaller at the origin.
Since meson decay constants are proportional to quark-antiquark wave
functions at the origin, decay constants in the valence approximation
will be smaller than in the full theory.

Thus while Ref. \cite{Butler93} confirms the valence approximation for
the low momentum behavior of the chromoelectic field governing the
binding energy of low-lying hadrons, the present calculation shows that
the valence approximation appears to become less reliable, as expected,
for the higher momenta contributing to meson wave functions at the
origin.  Our result suggests that valence approximation predictions for
the hadronic matrix elements entering weak decays will also tend to be
somewhat smaller than their values in the full theory.

The calculations described here were done on the GF11 parallel computer
at IBM Research \cite{Weingar90} and use the same collection of gauge
configurations and quark propagators generated for the mass calculations
of Ref. \cite{Butler93}.  The full set of mass and decay constant
calculations took approximately one year to complete.  GF11 was used in
configurations ranging from 384 to 480 processors, with sustained speeds
ranging from 5 Gflops to 7 Gflops.  With the present set of improved
algorithms and 480 processors, these calculations could be repeated in
less than four months.

The normalization we adopt for pseudoscalar and vector decay constants
in continuum QCD is $<\! 0 | J^{5 \mu}_j | P(p, j)\! > \; = i p^{\mu}
f_j $ and $<\! 0 | J^{\mu}_j | V(p, \epsilon, j)\! > \; = \epsilon^{\mu}
m_j F_j$, for states normalized by $ < \!p | q \!> \; = (2 \pi)^3 p^0
\delta( \vec{p} - \vec{q})$, where $J^{\mu}_j$ and $J^{5 \mu}_j$ are
vector and axial vector flavor-SU(3) currents and $j$ is an
adjoint-representation flavor index running from 1 to 8. Assuming exact
isospin symmetry, we have $f_i = f_{\pi}$, for $i = 1, ... 3$, $f_i =
f_K$, for $i = 4, ...7$, and $F_i = m_{\rho} / f_{\rho}$, for $i = 1,
... 3$. In the valence approximation we have, in addition, $F_8 = 3
m_{\phi} /(\sqrt{2} f_{\phi})$.  For simplicity, we will also use the
names $F_{\rho}$ and $F_{\phi}$ for $F_1$ and $F_8$, respectively. Our
normalization gives $f_{\pi}$ the value $(93.15 \pm 0.11)$MeV.

The hadron mass calculation of Ref. \cite{Butler93} was done using
gaussian smeared quark and antiquark fields defined in Coulomb gauge.
Smeared fields have, therefore, also been adopted for the present
calculation.  The smeared field $\phi_r(\vec{x},t)$ is $\sum_{\vec{y}}
G_r(\vec{x} - \vec{y}) \psi(\vec{y},t)$ for a smearing function
$G_r(\vec{z})$ of $(\sqrt{\pi}r)^{-3} exp( - |\vec{z}|^2 / r^2)$, and
$\overline{\phi}_r(\vec{x},t)$ is defined correspondingly.  We take the
smeared fields $\phi_0(x)$ and $\overline{\phi}_0(x)$ to be $\psi(x)$
and $\overline{\psi}(x)$, respectively. From these fields, define the
composite field $J^5_{j r}$ and smeared lattice currents $J^{\mu}_{j r}$
and $J^{5 \mu}_{j r}$ to be $ \overline{\phi}_r \gamma^5 \lambda_j
\phi_r$, $\overline{\phi}_r \gamma^{\mu} \lambda_j \phi_r$ and
$\overline{\phi}_r \gamma^5 \gamma^{\mu} \lambda_j \phi_r$,
respectively, where the $\lambda_j$ are an orthonormal basis for the
flavor-SU(3) Lie algebra with normalization $tr( \lambda_j \lambda_j) =
1/2$.

Define the constants $Z^{PP}_{j r' r}$, $Z^{AP}_{j r' r}$, and
$Z^{VV}_{j r' r}$ by the requirement that for a lattice with time
direction period L and a time $t$, $L \gg t \gg 1$, the correlation
function $\sum_{\vec{x}} <\! [J^5_{ j r'}(\vec{x},t)]^{\dagger} J^5_{j
r}(0,0)\!>$ approaches $Z^{PP}_{j r' r} exp( -m^P_j t)$, $\sum_{\vec{x}}
<\![J^{5 0}_{ j r'}(\vec{x},t)]^{\dagger} J^5_{j r}(0,0)\!>$ approaches
$Z^{AP}_{j r' r} exp( -m^P_j t)$, and $\sum_{\vec{x}} <\![J^i_{j
r'}(\vec{x},t)]^{\dagger} J^i_{ j r}(0,0)\!>$ approaches $Z^{VV}_{j r'
r} exp( -m^V_j t)$.  Here $m^P_j$ and $m^V_j$ are pseudoscalar and
vector masses, respectively.

Measured in units of the lattice spacing $a$ with any choice of the
smearing radius $r$, the decay constant $f_j a$ then is given by
$2^{\frac{1}{2}}( m_j a Z^{PP}_{jrr})^{-\frac{1}{2}} z^A_j Z^{AP}_{j 0
r} $, and $F_j a$ is given by $2^{\frac{1}{2}}( m_j a
Z^{VV}_{jrr})^{-\frac{1}{2}}z^V_j Z^{VV}_{j 0 r} $.  The coefficients
$z^A_j$ and $z^V_j$ are finite renormalizations chosen so that the
lattice currents $z^A_j J^{5 \mu}_{j r}$ and $z^V_j J^{\mu}_{j r}$
approach the continuum currents $J^{5 \mu}_j$ and $J^{\mu}_j$,
respectively, as the lattice spacing approaches zero.  A mean-field
theory improved perturbation expansion \cite{Lepage} gives $z^A_j$ the
value $(1 - 3 k_j / 4 k_c) [1 - 0.31 \alpha_{\overline{ms}}( 1 / a)]$
and $z^V_j$ the value $(1 - 3 k_j / 4 k_c) [1 - 0.82
\alpha_{\overline{ms}}( 1 / a)]$, where $k_j$ is the hopping constant
corresponding to the mass of the quark and antiquark for a meson with
flavor $j$, assumed here to have $m_q = m_{\overline{q}}$, and $k_c$ is
the critical hopping constant at which the pion's mass becomes zero.
The decay constant for a meson with $m_q \neq m_{\overline{q}}$ will be
discussed below.  The renormalization constants $z^A_j$ and $z^V_j$ are
often given the ``naive'' value $2 k_j$. Even to leading order in
perturbation theory, however, a better choice turns out to be $(1 - 3
k_j / 4 k_c)$ \cite{Lepage}.

Table~\ref{tab:lattices} lists the lattice sizes, parameter values,
sweeps skipped between gauge configurations, and number of
configurations used in the ensembles from which decay constants were
calculated. Gauge configurations were generated with the
Cabbibo-Marinari-Okawa algorithm.  A discussion of the algorithms by
which quark propagators were found is given in Ref. \cite{Butler93}.
Hadron masses were determined by fits to hadron propagators constructed
from the quark propagators.  The coefficients $Z^{AP}_{j 0 2}$,
$Z^{PP}_{j 2 2}$, $Z^{VV}_{j 0 2}$, and $Z^{VV}_{j 2 2}$ were then
extracted from fits to current-current and pseudoscalar-current
correlation functions, and combined with both perturbative and naive
renormalization to find two different sets of values for $f_{\pi}$ and
$F_{\rho}$.  Statistical errors on these quantities and all other
parameters which we have determined were found by the bootstrap method.
A more detailed discussion of our fits and error analysis will be given
elsewhere~\cite{Toappear}.

Percentage changes in decay constants going from $16^3 \times 32$ to
$24^3 \times 32$, at $\beta$ of 5.7, are given in
Table~\ref{tab:voldep}.  These changes are the same for both
perturbative and naive renormalization.  All of the differences appear
to be of marginal statistical significance and may therefore best be
viewed as upper bounds on the volume dependence of our results.  It
appears quite likely that for the range of $k$, $\beta$, and lattice
volume we have examined the errors in valence approximation decay
constants due to calculation in a finite volume $L^3$ are bounded by
an expression of the form $C e^{- L/R}$, with a coefficient $R$ of the
order of the radius of a hadron's wave function. At $\beta$ of 5.7 for
the $k$ we considered, $R$ is thus typically 3 lattice units or less.
We therefore expect that the differences between decay constants on a
$16^3$ volume and those on a $24^3$ volume are nearly equal to the
differences between $16^3$ and true infinite volume limiting values.

At the largest $k$ on each lattice, the ratio $m_{\pi} / m_{\rho}$ is
significantly larger than its experimentally observed value of 0.179.
Thus to produce decay constants for hadrons containing only light
quarks, our data has to be extrapolated to larger $k$ or, equivalently,
to smaller quark mass.  We did not calculate directly at larger $k$ both
because the algorithms we used to find quark propagators became too slow
and because the statistical errors we found in trial calculations became
too large.

To extrapolate $F_{\rho}$ and $f_{\pi}$ to small quark mass, we first
determined the critical hopping constant $k_c$, for each lattice and
$\beta$, at which $m_{\pi}$ becomes zero \cite{Butler93}.  Defining the
quark mass in lattice units $m_q a$ to be $1/(2 k) - 1/(2 k_c)$, we
found $f_{\pi} a$ and $F_{\rho} a$, both for perturbative
renormalization and for naive renormalization, to be nearly linear
functions of $m_q a$ over the entire range of $k$ considered on each
lattice.  Figure~\ref{fig:mextrap} shows $f_{\pi}$ and $F_{\rho}$, given
by perturbative renormalization, as functions of $m_q$.  Data is shown
from all lattices of Table~\ref{tab:lattices} except $24^3 \times 32$ at
$\beta = 5.7$.  For convenience, we show all hadron masses in units of
the physical rho mass, $m_{\rho}(m_n)$, given by $m_{\rho}$ evaluated at
the ``normal'' quark mass $m_n$ which produces the physical value of
$m_{\pi}/m_{\rho}$. The quark mass $m_q$ in Figure~\ref{fig:mextrap} is
shown in units of the strange quark mass $m_s$. The value of $m_s$ for
each lattice and $\beta$ is found in Ref. \cite{Butler93} by requiring
$m_{\pi}[ (m_n + m_s)/2] / m_{\rho}(m_n)$ to be equal to the physical
value of $m_K/m_{\rho}$.  The straight lines are fits to the $m_q$
dependence of $f_{\pi}$ and $F_{\rho}$ at the three smallest quark
masses in each data set at fixed $\beta$.  The fits in
Figure~\ref{fig:mextrap} appear to be quite good and provide, we
believe, a reliable method for extrapolating decay constants down to
light quark masses.  With naive renormalization, $f_{\pi}$ and
$F_{\rho}$ fit straight lines in $m_q$ about as well as the
perturbatively renormalized data of Figure~\ref{fig:mextrap}.

The linear fits of Figure~\ref{fig:mextrap} permit the determination of
$f_K$ and $F_{\phi}$ in addition to $f_{\pi}$ and $F_{\rho}$.  For a
pion composed of a quark and antiquark with mass $m_q \neq
m_{\overline{q}}$, Figure~\ref{fig:mextrap} suggests $f_{\pi} = \alpha_q
m_q + \alpha_{\overline{q}} m_{\overline{q}} + \beta$.  Charge
conjugation invariance then gives $\alpha_q = \alpha_{\overline{q}}$.
It follows that the kaon, which is a pion with, say, $m_q = m_s$ and
$m_{\overline{q}} = m_n$, will have the same decay constant as a pion
composed of a single type of quark and antiquark with $m_q =
m_{\overline{q}} = (m_s + m_n)/2$.  On the other hand, the linear fits
of Figure~\ref{fig:mextrap} permit $F_{\rho}$ to be extrapolated to the
point $m_q = m_{\overline{q}} = m_s$ which, in the valence
approximation, gives $F_{\phi}$.

The ratios $f_{\pi} / m_{\rho}$, $f_K / m_{\rho}$, $F_{\rho} / m_{\rho}$
and $F_{\phi} / m_{\rho}$ for physical quark masses we then extrapolated
to zero lattice spacing.  The physical lattice volume was held nearly
fixed as the lattice spacing was taken to zero.  For Wilson fermions the
leading lattice spacing dependence in mass ratios is expected to be
linear in $a$.  Figure~\ref{fig:aextrap} shows decays constants with
perturbative renormalization along with linear fits to $m_{\rho} a$. The
quantity $m_{\rho} a$ may be viewed as the lattice spacing $a$ measured
in units of the physical rho Compton wavelength, $1/m_{\rho}$.  The
vertical bars at $m_{\rho} a$ of 0, offset slightly for visibility, are
the extrapolated predictions' uncertainties determined by the bootstrap
method, and the horizontal bars are corresponding experimentally
observed values.  The calculated points in Figure~\ref{fig:aextrap} are
for the lattices $16^3 \times 32$, $24^3 \times 36$ and $30 \times 32^2
\times 40$.  The values of $\beta$ for these lattices were chosen so
that the physical volume in each case is nearly the same.  For lattice
period L, the quantity $m_{\rho} L$ is respectively, 9.08 $\pm$ 0.13,
9.24 $\pm$ 0.19 and, averaged over three directions, 8.67 $\pm$ 0.12
\cite{Butler93}.  Extrapolations
to zero lattice spacing similar to those shown in
Figure~\ref{fig:aextrap} were also done for naive renormalization.

The continuum ratios we found in finite volume were then corrected to
infinite volume by an adaptation of the method used in Ref.
\cite{Butler93} to correct finite volume continuum mass ratios to
infinite volume.  The difference between a unitless decay ratio
found on the lattice $16^3 \times 32$, at $\beta$
of 5.7, and the same ratio found on the lattice $24^3 \times 32$, at
$\beta$ of 5.7, we took as a finite lattice spacing approximation to the
difference between the continuum decay ratio in a box with period having
$m_{\rho} L$ of 9 and the continuum decay ratio in infinite volume.  The
error in this procedure, for perturbatively renormalized decay
constants, can be estimated to be 4 \% or less as follows.  For
perturbative renormalization, the value of each decay ratio on the
lattice $16^3 \times 32$ at $\beta$ of 5.7 differs from its continuum
limit by at most 75 \% of the continuum limit.  Moreover, as we argued
earlier, the change in each decay ratio, at $\beta$ of 5.7,
between $16^3 \times 32$ and $24^3 \times 32$ should be nearly the same
as the corresponding change between $16^3 \times 32$ and infinite volume.
Combining these two pieces of information, we expect that with a
relative error of 75 \% or less, the change in any decay ratio
between $16^3 \times 32$ and $24^3 \times 32$ at $\beta$ of 5.7 should
be the same as the change between the continuum decay ratio in a box with
period having $m_{\rho} L$ of 9 and the corresponding continuum ratio in
infinite volume.  Since the changes we found in decay ratios,
for physical quark mass, between $16^3 \times 32$ and $24^3
\times 32$ are all less than 5 \%, the error in using these
differences as estimates of corresponding continuum differences between
$m_{\rho} L$ of 9 and infinite volume should of the order of 75 \% of
5\%, which is 3.75 \%. The corresponding uncertainty for naive
renormalization is 10.0 \% due to a larger change in these values
between $16^3 \times 32$ at $\beta$ of 5.7 and the continuum limit.

The ratios $f_{\pi} / m_{\rho}$, $f_K / m_{\rho}$, $F_{\rho} / m_{\rho}$
and $F_{\phi} / m_{\rho}$, for both naive and perturbative
renormalization, extrapolated to zero lattice spacing with $m_{\rho} L$
fixed at 9, and then corrected to infinite volume are shown in
Table~\ref{tab:res}. The errors shown for infinite volume ratios are
statistical only and do not include the estimates we have just given for
the systematic error in our procedure for making infinite volume
corrections.  Most of the perturbatively renormalized results lie below
the corresponding ratios with naive renormalization.  For all ratios
except $F_{\phi} / m_{\rho}$ in finite volume, however, the two
different renormalizations give results which are statistically
consistant.  Where there is a disagreement, the perturbatively
renormalized ratios are more reliable estimates of the true infinite
volume continuum limit of the valence approximation.  For extrapolations
done from lattice spacings which are sufficiently small, the two
different renormalization procedures should give the same limiting
results. The agreement between the two different methods for three of
four parameters tends to support the reliability of our extrapolations
to the continuum limit.

The predicted infinite volume ratios in Table~\ref{tab:res} are
statistically consistent with the finite volume ratios. The main
consequence of the correction to infinite volume is an increase in the
size of the statistical uncertainty in each prediction.

Three of the four finite volume volume continuum ratios obtained by
perturbative renormalization lie below experiment by between one and
two standard deviations. The larger physical volume over which
infinite volume meson wave functions can spread should give infinite
volume decay constants some amount smaller than those calculated in
finite volume.  Thus the true disagreement between infinite volume
$f_{\pi} / m_{\rho}$, $f_K / m_{\rho}$, and $F_{\rho} / m_{\rho}$ and
experiment should be at least as great than the disagreement we find
for the finite volume predictions.  The statistical significance of
the disagreement between our inifinite volume results themselves and
experiment, however, is weaker than for the finite volume predictions
as a result of the large size of the error bars on the infinite volume
predictions.  The larger error bars of the naive renormalized decay
constants also weakens the significance of their disagreement with
experiment.

We would like to thank Paul Mackenzie for discussions, and Mike Cassera,
Molly Elliott, Dave George, Chi Chai Huang and Ed Nowicki for their work
on GF11.

\figure{ The decay constants $F_{\rho}$ and $f_{\pi}$, with perturbative
renormalization, in units of the physical rho mass $m_{\rho}(m_n)$, as
functions of the quark mass $m_q$, in units of the strange quark mass
$m_s$.
\label{fig:mextrap}}

\figure{ Perturbatively renormalized
decay constants as functions of the lattice
spacing $a$, in units of $1/m_{\rho}$. The error bars near zero lattice
spacing are uncertainties in the extrapolated ratios, and the horizontal
lines represent experimentally observed values.
\label{fig:aextrap}}

\begin{table}
\begin{tabular}{r@{}llllr}     \hline
 \multicolumn{2}{c}{lattice}          & $\beta$ & k     & skip & count\\
 \hline
 $16^3$ & $\:\times \: 32$ & 5.7 & 0.1600 - 0.1675 & 2000 & 219\\
 $24^3$ & $\:\times \: 32$ & 5.7 & 0.1600 - 0.1675 & 4000 & 58\\
 $24^3$ & $\:\times \: 36$ & 5.93 & 0.1543 - 0.1581 & 4000 & 217\\
 $30 \times 32^2$ & $\:\times \: 40$ & 6.17 & 0.1500 - 0.1532 & 6000 & 219\\
 \hline
\end{tabular}
\caption{Configurations analyzed.}
\label{tab:lattices}
\end{table}

\begin{table}
\begin{tabular}{crr}     \hline
 decay   & k & change \\ \hline
$f_{\pi}$ & 0.1600 &  $ 3.5^{+3.2}_{-2.3} \% $ \\
          & 0.1650 &  $-1.1^{+3.3}_{-5.4} \% $ \\
          & 0.1663 &  $-3.1^{+4.1}_{-5.0} \% $ \\
          & 0.1675 &  $-6.2^{+4.7}_{-7.1} \% $ \\ \hline
$f_{\rho}$ & 0.1600 &  $ 7.1^{+1.4}_{-4.9} \% $ \\
          & 0.1650 &   $ 4.9^{+3.0}_{-7.8} \% $ \\
          & 0.1663 &   $ 4.0^{+1.6}_{-7.8} \% $ \\
          & 0.1675 &   $-2.3^{+1.6}_{-3.9} \% $ \\ \hline
\end{tabular}
\caption{
Changes in perturbatively renormalized decay constants from $16^3 \times
32$ to $24^3 \times 32$ at $\beta = 5.7$}
\label{tab:voldep}
\end{table}

\begin{table}
\begin{tabular}{cllll}     \hline
 decay  & renorm. & finite volume     & infinite volume  & obs. \\ \hline
$f_{\pi} / m_{\rho}$ & perturb.  & $0.102^{+0.021}_{-0.026}$ &
                                   $0.091^{+0.028}_{-0.036}$ & 0.121 \\
                     & naive     & $0.097^{+0.030}_{-0.032}$ &
                                   $0.082^{+0.041}_{-0.051}$ & \\ \hline
$f_K / m_{\rho}$     & perturb.  & $0.119^{+0.017}_{-0.015}$ &
                                   $0.116^{+0.021}_{-0.024}$ & 0.148 \\
                     & naive     & $0.130^{+0.024}_{-0.022}$ &
                                   $0.127^{+0.030}_{-0.032}$ & \\ \hline
$F_{\rho}/ m_{\rho}$ & perturb.  & $0.177^{+0.017}_{-0.020}$ &
                                   $0.173^{+0.039}_{-0.031}$ & 0.199 \\
                     & naive     & $0.191^{+0.033}_{-0.031}$ &
                                   $0.201^{+0.035}_{-0.059}$ & \\ \hline
$F_{\phi}/ m_{\rho}$ & perturb.  & $0.217^{+0.014}_{-0.022}$ &
                                   $0.253^{+0.017}_{-0.051}$ & 0.219 \\
                     & naive     & $0.270^{+0.021}_{-0.029}$ &
                                   $0.319^{+0.030}_{-0.070}$ &  \\ \hline
\end{tabular}
\caption{Calculated values of meson decay constants
extrapolated to zero lattice spacing in finite volume,
then corrected to infinite volume, compared with observed
values.}
\label{tab:res}
\end{table}

\end{document}